\documentclass[twocolumn,showpacs,preprintnumbers]{revtex4}%
\usepackage{graphicx}
\usepackage{dcolumn}
\usepackage{bm}%
\usepackage{amsmath}%
\usepackage{amsfonts}%
\usepackage{amssymb}

\newcommand{\be}{\begin{equation}}
\newcommand{\ee}{\end{equation}}
\newcommand{\bea}{\begin{eqnarray}}
\newcommand{\eea}{\end{eqnarray}}

\begin{document}
\title{ Pair Production in Non-Perturbative QCD }
\author{S. Hamieh}
\affiliation{\textit{Department of Physics,  Faculty of Sciences, Lebanese University,
Beirut Lebanon.}}
\date{\today}

\begin{abstract}
In this note, we present a method to calculate the  vacuum to vacuum transition
amplitude in the
presence of the non-abelian background field.  The number of non-perturbative  quark-antiquark produced per unit time,  per unit volume and per unit transverse momentum from a given constant chromo-electric field is calculated and compared with
the results found in the literature.
\end{abstract}
\pacs{03.67.Hk}
\maketitle

Lattice QCD predict a phase transition from Hadrons
gaz (HG) to quark-gluon plasma (QGP) at deconfinement temperature, T $\sim$ 170MeV.
 It is believed that QGP
has been produced in  relativistic heavy ions collision \cite{Hami03,ham4}.
In the initial pre-equilibrium stage of QGP about
half the total center-of-mass energy, $E_{cm}$, goes into the
production of a semi-classical gluon field~\cite{all,all1}.
To study the
production of a QGP from a classical chromo field,
it is necessary to know how quarks and gluons are formed
from the latter.
The production rate of
 quark-antiquark  from a given constant chromo-electric field $E^a$  has been derived in  Ref. \cite{paper1}
 and the integrated $p_T$ distribution has been
obtained  in \cite{schw,yildiz1,wkb1,wkb2} (for a review see \cite{dune}). The $p_T$ distribution for quark (antiquark) production
can be used in the analysis of the experimental results at the RHIC and the LHC colliders.
In this short technical note, we will extend the results of Ref. \cite{paper1} to a general constant background field using a different method. The method presented here may simplify the complexity
 found in the Non-perturbative QCD calculations.

For this purpose, we start from the QCD Lagrangian density for a quark in a non-abelian background field
$A_{\mu}^a$ which is given by
\bea
{\cal{L}}~
=~\bar{\psi}~ [(~\hat{{p}\!\!\!\slash}
~-~gT^a{{A}\!\!\!\slash}^a) -m]
~\psi ~=~\bar{\psi}~ D[A] ~\psi\,,
\label{laq}
\eea
Then the vacuum to vacuum transition
amplitude is given by
\bea
 \langle 0|0\rangle~&=&~\frac{\int~{\cal{D}}\bar{\psi}{\cal{D}}\psi~
e^{i\int d^4x~\bar{\psi}~D[A]~\psi}}{
\int~{\cal{D}}\bar{\psi}{\cal{D}}\psi~ e^{i\int d^4x~\bar{\psi}~D[0]~\psi}}\nonumber\\
~ &=&~{\rm Det}[D[A]]/{\rm Det}[D[0]] \,.
\eea
And the one loop effective action can be written in this form
\bea
S~=~-i\ln\langle 0|0\rangle~~=~ -i~{\rm Tr}\,\ln
[\frac{(~\hat{{p}\!\!\!\slash}
~-~gT^a{{A}\!\!\!\slash}^a) -m}{~\hat{{p}\!\!\!\slash}  -m}]\,.
\label{efgl}
\eea
Using the invariance of trace under transposition and the following relation
 \bea
\ln\frac{a}{b}~=~\int_0^\infty~\frac{ds}{s} [
e^{is~(b~+i\epsilon)}~ -e^{is~(a~+i\epsilon)}]\,,
\label{eone}
\eea
we obtain   the following expression
\footnote{see Ref. \cite{paper1} and reference therein}
%\begin{widetext}
\bea
2S&=&i{\rm Tr} \int_0^\infty\frac{ds}{s}\Big[\exp is
[(\hat{{p}} -gT^a{{A}}^a)^2 +
\frac{g}{2}\sigma_{\mu \nu}T^aF^{a \mu \nu}\nonumber\\&&
-m^2+i\epsilon] - \exp is [~\hat{{p}}^2  -m^2+i\epsilon]\Big]\,.
\label{trhg}
\eea
%\end{widetext}
The
quickest way to calculate the effective action is to work in a basis $|\Psi\rangle$ of eigenstates of
\be~\hat{H}=(~\hat{{p}} ~-~gT^a{{A}}^a)^2 ~+~
\frac{g}{2}\sigma_{\mu \nu}T^aF^{a \mu \nu}\,.\ee
%Since we are interested by the evaluation of the vacuum to vacuum transition amplitude due to constant background field,
Now, we consider first the case of  a constant electric field in the $z$ direction (direction of the beam in the heavy ion collision). In this case, we choose a gauge such that we can take
$A^a_z =E^at$. Thus
\bea\frac{g}{2}\sigma_{\mu \nu}T^aF^{a \mu \nu}=i~g~E^a~T^a~\sigma_3\otimes\left( \begin{array}{cccc}
 0& 1 \\
1 & 0
      \end{array} \right)\eea
 So the Hamiltonian becomes
\bea~\hat{H}&=&~\hat{{p_t}}^2 ~-~\hat{{p_x}}~^2 ~-~\hat{{p_y}}^2 \nonumber\\&~-~&(\hat{{p_z}}~-~gT^aE^at)^2 ~+~
\frac{g}{2}\sigma_{\mu \nu}T^aF^{a \mu \nu}
\eea
 The eigenvalue of this Hamiltonian are

\bea E_n^{p_x,p_y,p_z,\Lambda_i,\lambda_j}=\hat{{-p_T}}^2 - g\lambda_j(2n+1) ~+~
{ig}\Lambda_i\lambda_j
\,.
\label{trg}
\eea
Where $\Lambda_i$ are the eigenvalues over the Dirac matrices such that
$\Lambda_1=\Lambda_3=1,\, $ and $\Lambda_2=\Lambda_4=-1$.
And $\lambda_j$, with $j=1,2,3$, are the eigenvalue for $\lambda=T^aE^a$ over the group space and  given in Ref. \cite{paper1}.
 Using these eigenvalues the effective action becomes
 %we can now proceed in the evaluation o $S$ as follow
 \begin{widetext}
 \bea
2S~&=&~i~\int_0^\infty~\frac{ds}{s}~\sum_{i=1}^4~\sum_{j=1}^3
~ \frac{1}{(2\pi)^3}~\int d^4x~\int
d^2p_T~
e^{-is(p_T^2~+~m^2) ~-~s\epsilon}\Big[~\sum_{n=0}^{\infty}~|g\lambda_j|
e^{ sg\lambda_j(2n+1) ~-~
{sg}\Lambda_i\lambda_j}-\frac{1}{2s}~\Big]\,.
\label{trg}
\eea
 \end{widetext}
 Performing the $i$ and $n$ summations we found
\bea
2S~&=&~i~\int_0^\infty~\frac{ds}{s}~\sum_{j=1}^3
~ \frac{1}{4\pi^3}~\int d^4x~\int
d^2p_T~\nonumber\\&&
e^{-is(p_T^2~+~m^2) -s\epsilon}[|g\lambda_j|
\frac{\cosh sg\lambda_j}{\sinh s|g\lambda_j|}-\frac{1}{s}~]\,,
\label{trg}
\eea
which is the same results of Ref. \cite{paper1}. Clearly,  the one loop magnetic effective action can be found upon the following substitution $E^a\rightarrow -i~B^a$. Therefore
\bea
2S^{(m)}~&=&~i~\int_0^\infty~\frac{ds}{s}~\sum_{j=1}^3
~ \frac{1}{4\pi^3}~\int d^4x~\int
d^2p_T~\nonumber\\&&
e^{-is(p_T^2~+~m^2) -s\epsilon}[|g\lambda_j|
\frac{\cos sg\lambda_j}{\sin s|g\lambda_j|}-\frac{1}{s}]\,,
\label{trg}
\eea
%where the $\lambda_j$ are the eigenvalues of the matrix $\lambda=T^aB^a$.
Now, in the same manner as in Ref. \cite{paper1} we may derive the non-perturbative quarks
(antiquarks) production per unit time, per unit volume and per unit transverse
momentum from a given constant chromo-electric field $E^a$
\begin{figure}[t!]
   \centering
   \includegraphics[width=0.9\columnwidth]{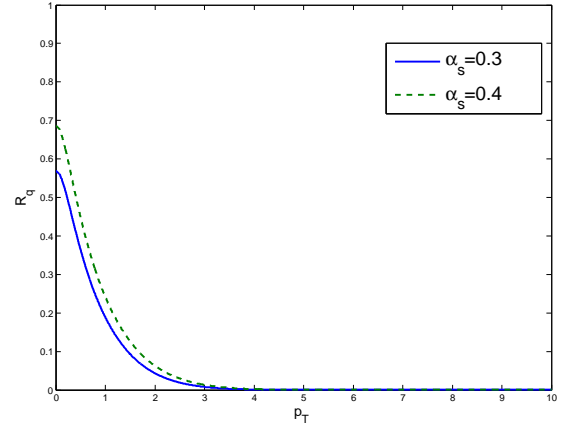}
   \caption{\label{fig3}(Color online)
   Transverse production rate for quarks for $C_1 = 100$  for $\alpha_s=0.3, 0.4$, as a function of~$p_\mathrm{T}$.
   For simplicity we denote here the quark production rate given in Eq.~\eqref{2} by $R_q$. We take $\theta=0$, $m=m_q \approx 1/3$~GeV.}
\end{figure}
\begin{align}
   &
   \frac{\mathrm{d}N_{q,\bar q}}
     {\mathrm{d}t \, \mathrm{d}^3x \, \mathrm{d}^2 p_\mathrm{T}}
   \label{2} \\
   &=
   -
   \frac{1}{4\pi^3} \, \sum_{j=1}^3 \, |g\lambda_j| \,
   \ln
   \Bigl \{ \,
      1
      -
      \exp
      \Bigl [
         - \frac{ \pi ( p_T^2+m^2 )}{|g\lambda_j|}
      \Bigr ] \,
   \Bigr \}  \>,
   \notag
\end{align}
where $m$ is the effective mass of the quark and the eigenvalues $\lambda_j$ are given by
\begin{align}
   & \lambda_1=\sqrt{\frac{C_1}{3}}~\cos \theta\,,
   \notag \\
   & \lambda_2=\sqrt{\frac{C_1}{3}}~\cos (2\pi/3-\theta)\,,
   \notag \\
   & \lambda_3=\sqrt{\frac{C_1}{3}}~\cos (2\pi/3+\theta)
   \>,
\end{align}
with $\theta$ given by
\begin{align}
   \cos^2 (3 \theta ) = 3 \, C_2/C_1^3
   \>.
\end{align}
where
\begin{align}
   C_1=E^aE^a \>,
   \qquad
   C_2=[d_{abc}E^aE^bE^c]^2
   \>,
\end{align}
 Note that $0 \le C_1^3/(3C_2) \le 1$.

 As an application of the above results let's consider the situation of two relativistic heavy nuclei colliding and leaving behind a semi-classical gluon field which then non-perturbatively produces gluon and quark-antiquark pairs via the Schwinger mechanism~\cite{schw}.
 As estimated in Ref. \cite{coop}   for Au-Au collision at RHIC collider with $R \approx 10$~fm and  center-of-mass energy  $\approx 200$~GeV per nucleon,  the initial energy density is  $\rho \approx  100$~GeV$^4$ and $C_1 \sim 100$ ~GeV$^4$. For our analysis we take $\theta=0$ which can be justified by the sensitivity check that has been made in Ref. \cite{coop} where it has been  found that the production rate is not
  very sensitive to $C_2$.

In Fig.~\ref{fig3} we plot the rate of quark production as a function of the transverse momentum for two values of   $\alpha_s$ = $0.3$ used in \cite{aur}  and $\alpha_s=0.4$ for  the initial energy density of $\rho \approx  100$~GeV$^4$.

In conclusion, in this note we have proposed a method for calculating the  vacuum to vacuum transition
amplitude in the
presence of the non-abelian background field.  The method can be applied  to
a general background field and it can be updated to study the non-perturbative soft gluon production \cite{paper2}.
Also, we have evaluated the  rate for
 quark (antiquark) production in a constant chromo-electric field $E^a$.

\end{document}